\documentclass{article}
\usepackage{amsmath}
\usepackage{amssymb}
\usepackage{bm}
\usepackage{color}
\begin{document}
\begin{center}
\LARGE{Dirac's Equation in Different Numerical Rings}
\end{center}
\medskip
\begin{center}
Lester C. Welch

Aiken, SC 29803

lester.welch@gmail.com
\end{center}
\begin{quote}
By confining the elements of the matrices obtained in the linearization of a wave equation (Dirac's equation) to a particular mathematical ring, either $\mathbb{R}$, $\mathbb{C}$, or $\mathbb{H}$ \footnote{A field is a commutative division ring. A ring is an algebraic structure which generalizes the algebraic properties of the integers and contains two operations usually called addition and multiplication. One example of a field is the familiar complex numbers. A division ring allows for division (except by zero).  Every field is a ring but non-commutative rings are not fields.} and by searching for symmetries resulting from operators in each ring the three quantum mechanical systems are examined for any ability to reveal new physics.
\end{quote}
\section{Introduction and History \normalsize{ \cite{schiff,messiah}}}

Dirac, \cite{dirac} using $\mathbb{C}$, derived a first order equation  linear in momentum and energy, describing a relativistic quantum mechanical particle starting with:
\begin{equation}
(A\partial_x + B\partial_y + C\partial_z + iD\partial_t - m)\psi = 0\label{eqn:d0}
\end{equation}
which can be rewritten as:
\begin{equation}
\gamma_\mu \partial_\mu\psi + m\psi = 0 \label{eqn:d1}
\end{equation}where
\begin{equation} x_\mu = (x,y,z,\tau) \mbox{  and  }  \tau = it\mbox{        (}\mu\mbox{= 1,...,4)}\label{eqn:it}
\end{equation}In order to recover  
\begin{equation}
(\nabla^2 - \frac{1}{c^2}\;\partial^2_t)\psi =  m^2\psi \label{eqn:kg}
\end{equation} the Klein-Gordon equation, it is necessary that 
\begin{equation}
\gamma_\mu\gamma_\nu + \gamma_\nu\gamma_\mu = 2 \delta_{\mu\nu} \cdot \textbf{I} \label{eqn:com}
\end{equation}These relationships cannot be fulfilled by scalars and thus, matrices are used.  By choosing, $\mathbb{C}$, to create the elements of the required matrices, at least 4x4 matrices are needed to achieve the required commutation properties ($\ref{eqn:com}$) - so the wave function had four components.  At the time of Dirac's derivation Pauli had discovered a non-relativistic theory, involving 2x2 matrices - defined as the Pauli spinors - to explain the electron's behavior in magnetic fields.  Thus Dirac's theory had twice as many components to the wave function as Pauli's.  Dirac found a representation of the 4x4 matrices, called the Dirac matrices, that used the Pauli spinors as elements - for historical reasons. When Dirac's equation is made guage invariant, both globally and locally, it leads to the extremely successful theory of quantum electrodynamics.  This paper seeks to examine the scope of Dirac's equation by confining the elements of Dirac's matrices to a particular ring.
\section{ Significance of the Choice of the Mathematical Ring}

It is well known (see, for example, \cite{Adler}) that only three rings, $\mathbb{R}$, (the "reals"), $\mathbb{C}$, (the "complexes") and $\mathbb{H}$, (the quaternions, see Appendix A) are candidates for constructing quantum mechanical descriptions, call them QM$_R$, QM$_C$, and QM$_H$, of physical systems. It will be assumed that physical systems are embedded in a Minkowski space with a signature that has one negative, e.g., (-1,1,1,1)  The existence of the negative coefficient of the time coordinate means, that in the ring used to construct a mathematical description of the physical system, a square root of -1, \textbf{I}, must be represented so that invariant norms (e.g., $x^2 + y^2 + z^2 - t^2$) can be constructed.  Most familiar is the quantity $i$ in $\mathbb{C}$ which gives rise to complex numbers. 

There is no scalar quantity in $\mathbb{R}$, $\textbf{I}_R$, for which ($\textbf{I}_R)^2 = -1$, thus a more complicated structure (e.g., a matrix) must be constructed over $\mathbb{R}$. For example let\footnote{The matrix given is a specific example of the most general case in $\mathbb{R}$ of $\left(\begin{array}{cc} tan(\alpha) &  -sec(\alpha) \\
                       sec(\alpha) &  -tan(\alpha)
      \end{array}\right)$ where $\alpha\in\mathbb{R}$ is an arbitrary parameter akin to "phase" in $\mathbb{C}$.},
\[
\textbf{I}_R =\left(\begin{array}{cc} 0 &  1 \\
                       -1 &  0
      \end{array}\right);\hspace{.2in}  \textbf{I}_R \cdot \textbf{I}_R =
\left(\begin{array}{cc} -1 &  0 \\
                        0 &  -1
      \end{array}\right) = -1 \otimes \textbf{I}_2.
\]  This necessarily requires that the wave function, $\psi$, in QM$_R$, have at least two components and that $\textbf{I}_R$ will, in general, mix those components.  In the simplest illustrative case
\[\textbf{I}_R\psi = \left(\begin{array}{cc} 0 &  1 \\
                       -1 &  0
      \end{array}\right)\cdot
      \left(\begin{array}{c}\psi_1 \\ \psi_2 \end{array}\right) =
       \left(\begin{array}{c}\psi_2 \\ -\psi_1 \end{array}\right)
\]
This operation is analogous to the operation, in $\mathbb{C}$, of multiplying a complex number by $-\mathbf{I}_C \equiv -i$.   Hence the two components of the wave function in QM$_R$ can be viewed as isomorphic to a complex element within $\mathbb{C}$ - and thus associate $\psi_1$ with $\psi_R$ and $\psi_2$ with $\psi_C$. Since $\textbf{I}_R$ is not a scalar, it, in general, will not commute with other quantities.
 
In $\mathbb{C}$ the scalar quantity, $\textbf{I}_C$, commutes with $x \in \mathbb{C}$, so $\textbf{I}_C\psi = \psi \textbf{I}_C$.  In the quaternion ring, there are three independent scalar roots, $\textbf{I}_{Q:i,j,k}$ of -1 which do not commute.  So $\textbf{I}_Q\psi \neq \psi \textbf{I}_Q.$

\noindent To summarize:
\newline
\newline
\indent In $\mathbb{R}$, $\textbf{I}$ is not a scalar and mixes wave function components.
\newline
\indent In $\mathbb{C}$, $\textbf{I}$ is a scalar and commutes with the wave function.
\newline
\indent In $\mathbb{H}$, $\textbf{I}$ is a scalar and does not commute with the wave function.
\medskip
\newline
It should also be noted that, by using matices, any finite field with a subfield can be represented by that subfield.  One example, since $\mathbb{R}<\mathbb{C}$, is the representation of i ($\in \mathbb{C}$) by the rank two matrix, $\mathbf{I}_R$, containing only real elements.  Another example is the representation of $i,\;j,\;k\; (\in\mathbb{H})$ by rank two matrices containing the element $i = \mathbf{I}_C$ from $\mathbb{C}$ as follows:
\[ i_H = \left(\begin{array}{cc}0&1\\-1&0\end{array}\right),\;  
j_H = \left(\begin{array}{cc}-i&0\\0&i\end{array}\right),\;
k_H = \left(\begin{array}{cc}0&i\\i&0\end{array}\right) 
\] which means, in principle, that it is possible to construct all of QM$_C$ and QM$_H$ by using $\mathbb{R}$ with matrices of sufficient rank.  The only advantage of using higher order fields is for economy of notation and because certain operations (e.g., "conjugation") are more obvious.
\medskip
\newline
For clarity, to avoid the explicit use of $\textbf{I}_C$, equation ($\ref{eqn:d0}$), from QM$_C$ is rewritten,  in the most general linearization as
\begin{equation}
\left(\gamma_t \partial_t +\gamma_x \partial_x + \gamma_y \partial_y + \gamma_z \partial_z \right)\psi \equiv \gamma_\mu\partial_\mu\psi =m\psi.  \label{eqn:dw}
\end{equation}
Where the coefficients, $\gamma_\mu$, have as elements, members of the mathematical ring being used.
To recover equation ($\ref{eqn:kg}\;$)(setting $c=1$), the following conditions must hold, true for whatever numerical field, and making no assumptions about the mathematical structure of the $\gamma_\mu$:
\begin{equation}\gamma_0\gamma_0=-1;\;\;\gamma_x\gamma_x=\gamma_y\gamma_y=\gamma_z\gamma_z=1;\;\;\gamma_\mu\gamma_\nu+\gamma_\nu\gamma_\mu=0,\;\; \text{if}\;\mu\ne\nu. \label{eqn:cnd}
\end{equation}
\section{Dirac's Equation in $\mathbb{C}$, QM$_C$}
We start with $\mathbb{C}$, QM$_C$, the most familar, to show the consistency of ($\ref{eqn:d0}$) with (\ref{eqn:dw}) - a linearized, first order equation devoid of any scalar $\not\in \mathbb{R}$ of a specific numerical field (e.g., $\mathbf{I}_C$=i).
\medskip
\newline
In $\mathbb{C}$, to meet the commutation relationships, required by (\ref{eqn:cnd}), rank 4 matrices are required which has $\textbf{I}_C$ (a scalar) in some elements, and by using the example for $\gamma_\mu\;$, given in Appendix B, it is easily seen that 
\begin{equation}
\left( \begin{array}{cc}
       {(\mathbf{\sigma_t}\partial_t-m}) & -i\vec{\mathbf{\sigma}}\cdot \mathbf{\nabla} \\
      i\vec{\mathbf{\sigma}}\cdot \mathbf{\nabla} & {-(\mathbf{\sigma_t}\partial_t+m)}
              \end{array}\right)
         \left( \begin{array}{c}
     \Psi_1\\
     \Psi_2 
     \end{array}\right)= 0  \label{eqn:Ceq}            
\end{equation}
where
\[\Psi_1 = \left(\begin{array}{c}
           \psi_1\\
           \psi_2
           \end{array}\right);\hspace{.2in} \Psi_2=
           \left(\begin{array}{c}
           \psi_3\\
           \psi_4
           \end{array}\right)
\]
By setting
\[ \vec{p} = -i\nabla, \hspace{.2in} \partial_t = iE\mbox{ and noting that } i\sigma_t = -\mathbf{1}_2 = \left(\begin{smallmatrix} -1&0\\0&-1\end{smallmatrix}\right)\]
equation($\ref{eqn:Ceq}$) can be rewritten
\begin{equation*}
\left( \begin{array}{cc}
       -E-m & \vec{\mathbf{\sigma}}\cdot \vec{\mathbf{p}} \\
      -\vec{\mathbf{\sigma}}\cdot \vec{\mathbf{p}} & E-m
              \end{array}\right)
         \left( \begin{array}{c}
     \Psi_1\\
     \Psi_2 
     \end{array}\right)= 0            
\end{equation*}
or

\[\vec{\sigma}\cdot \vec{p}\;\Psi_2 - (m+E)\Psi_1 =0\]
\[\vec{\sigma}\cdot \vec{p}\;\Psi_1 + (m-E)\Psi_2 =0\]
which is identical\footnote{See for example, with minor differences in notation, equation 2.22 of Leon \cite{Leon.73}($\eta=\Psi_1\;\; \chi=\Psi_2$) or equation XX.180 of Messiah \cite{messiah}($\chi=\Psi_1;\;\Phi=\Psi_2$), the latter in the absence of any electromagnetic fields.} to the canonical treatment of Dirac's equation and thus equation ($\ref{eqn:dw}$) yields the same results as (\ref{eqn:d0}).
\section{Dirac's Equation in $\mathbb{R}$, QM$_R$}
By making the following replacements in the $\gamma_\mu$ in Appendix B: 
\begin{align*}
i\rightarrow \left(\begin{smallmatrix} 0 &  1 \\
-1 &  0\end{smallmatrix}\right),\quad\text{and}\quad 1 \rightarrow \left(\begin{smallmatrix} 1 &  0 \\
0 &  1\end{smallmatrix}\right)
\end{align*} 
the rank 8 $\gamma_\mu$ in Appendix C are obtained with elements $\in \mathbb{R}$.  The resulting solutions have twice as many components as in QM$_C$ but the physics in QM$_R$ from using $\left(\begin{smallmatrix} \psi_R \\
\psi_C \end{smallmatrix}\right)$ is isomorphic to $\psi=\psi_R + i \psi_C$ in QM$_C$.
There is a representation of the $\gamma_{\mu}$ which does not make explicit use of an imaginary - the Majorana representation\footnote{The Majorana representation should not be confused with the Majorana equation (an alternative to Dirac's equation) which leads to different physics, e.g. neutrinos which are their own anti-particle.} as given in Appendix D which differ from the $\gamma^C$ by a similiarity transformation and thus offer no new physics. For example:
\begin{align*}
S\gamma_\mu^CS^{-1} = \gamma_\mu^R\quad\text{ where}\quad S=\frac{1}{2}\left(\begin{array}{cccc} 1&i&1&-1\\1&-i&-1&-i\\i&1&i&-1\\i&-1&-1&-1\end{array}\right)
\end{align*}
Thus QM$_R$ offers nothing different from QM$_C$.

\section{Dirac's Equation in $\mathbb{H}$, QM$_H$}

Adler $\cite{Adler}$ has written an excellent book giving a comprehensive treatment and review of QM$_H$ but a difference in approach between Adler and the current work is that Adler's aim seems to be to supplant QM$_C$ with QM$_H$ He shows that asymptotically QM$_H$ and QM$_C$ give the same result and hopes that QM$_H$ will explain some of the details (e.g., "flavor" and "color") for which QM$_C$ is insufficient.   The perspective of the current work is that QM$_C$ does an excellent job of explaining its domain of physical reality and that, indeed, quantum electrodynamics, is the most accurate physical theory known.  Thus the ability of QM$_H$ to explain "old" physics (explained by QM$_C$) is not of concern, but that "new" (i.e., unexplained phenomenon) should be the focus of QM$_H$. 

Starting with ($\ref{eqn:dw}$), where it is to be understood that the elements of $\gamma_\mu$ and other numbers can now be from the quaternion ring $\mathbb{H}$, to satisfy ($\ref{eqn:cnd}$) four anticommuting quantities are needed.  Scalar quaternions have only three and thus can not be used.  Matrices of rank 2 are impossible since there is no antiunitary Hermitian matrix.  Thus rank 4 matrices must be used and one choice for the $\gamma_\mu$ is given in Appendix E.\footnote{It is easy to show that rank 3 matrices, $\mathbf{M}$, are incapable of satisfying ($\ref{eqn:cnd}$) by simply considering the fate of the central element, $m_{22}$.} The $\gamma_\mu$ also manifest a symmetry principle that each of $\emph{i,\;j,\;k\;}$ must have equal roles in that there is no reason, a priori, to favor one or the other.
The three complex units (\emph{i,\; j,\; k}) in $\mathbb{H}$ as compared to the single (\emph{i}) in $\mathbb{C}$ means that "complex conjugation" has to be clearly specified.  The following "modes" of conjugation are defined:
\[\Gamma_i (q) \equiv q^{*i} = a - b i + c j + d k,\mbox{     i-conjugation}\]
and similarly for $\Gamma_j$ and $\Gamma_k$.
\[\Gamma_{ij} (q) \equiv q^{*ij} = a - b i - c j + d k, \mbox{     ij conjugation}\]
and similarly for $\Gamma_{ik}$ and $\Gamma_{jk}$.
\[\Gamma_{ijk}(q) \equiv q^* = a -b i - c j - d k,\mbox{  triple (complete) conjugation}\]
It is clear that $\Gamma_i, \Gamma_j, \Gamma_k, \Gamma_{ij}$, etc., all commute.  It should also be noted that 
\[q\Gamma_{ijk}(q) \equiv qq^* = a^2 - b^2 - c^2 - d^2 = q^*q\mbox{  and} \in \mathbb{R}\]
\subsection{Single Quaternion Conjugation}
Suppose
\begin{equation*}
\psi = \left(\begin{array}{c} \psi_1 \\ \psi_2\\ \psi_3 \\ \psi_4\end{array}\right)
\end{equation*}
We define a transpose and $i$-conjugate, of $\psi$
\[\psi^{\dag i} = (\Gamma_i(\psi_1),\Gamma_i(\psi_2),\Gamma_i(\psi_3),\Gamma_i(\psi_4))\]
Taking the transpose and $i$-conjugate   of ($\ref{eqn:dw}$), results in
\begin{equation}
\partial_\mu\psi^{\dag i}(\gamma_\mu)^{\dag i} - m\psi^{\dag i} = 0 \label{eqn:d5}
\end{equation}
Further
\[(\gamma_\mu)^{\dag i} = (\gamma_0, \gamma_1, \gamma_2, \gamma_3)^{\dag i}=  (-\gamma_0, \gamma_1, -\gamma_2, -\gamma_3) \]
one can multiply on the right by $\gamma_3$ and taking advantage of the commutation relationships ($\ref{eqn:cnd}$)
\[(\gamma_\mu)^{\dag i}\gamma_3 =\gamma_3 (\gamma_0,-\gamma_1, \gamma_2, -\gamma_3)\]
and now multiply on the right by $\gamma_2$ to obtain
\[(\gamma_\mu)^{\dag i}\gamma_3\gamma_2 =\gamma_3 \gamma_2(-\gamma_0,\gamma_1, \gamma_2, \gamma_3)\]
and finally multiply by $\gamma_0$ to get
\[(\gamma_\mu)^{\dag i}\gamma_3\gamma_2\gamma_0 =\gamma_3\gamma_2\gamma_0 (-\gamma_0,-\gamma_1, -\gamma_2, -\gamma_3)\]
thus
\[(\gamma_\mu)^{\dag i}\gamma_3\gamma_2\gamma_0 = -\gamma_3\gamma_2\gamma_0\gamma_{\mu}\]
and ($\ref{eqn:d5}$) becomes
\begin{equation*}
\partial_\mu\psi^{\dag i}\gamma_3\gamma_2\gamma_0\gamma_\mu +m\psi^{\dag i}\gamma_3\gamma_2\gamma_0 = 0 
\end{equation*}
and to simplify notation
\[\psi^{\dag i}\gamma_3\gamma_2\gamma_0 = \bar{\psi^i} \] becomes
\begin{equation}
\partial_\mu\bar{\psi^i}\gamma_\mu +m\bar{\psi^i} = 0  \label{eqn:d7}
\end{equation}
Multiplying ($\ref{eqn:d7}$) on the right by $\psi$ and ($\ref{eqn:dw}$) on the left by $\bar{\psi^i}$ and adding gives:
\begin{equation}
\textcolor{red}{\partial_\mu(\bar{\psi^i}\gamma_\mu\psi)  = 0.}
\end{equation}
Likewise
\begin{equation*}
\textcolor{green}{\partial_\mu(\bar{\psi^j}\gamma_\mu\psi)  = 0.}
\end{equation*}
\begin{equation*}
\textcolor{blue}{\partial_\mu(\bar{\psi^k}\gamma_\mu\psi)  = 0.} 
\end{equation*}
\subsection{Double Quaternion Conjugation}
Double quaternion conjugation has the feature that
\[\Gamma_{jk}(\psi) = \Gamma_i(\psi^*)\mbox{,   or}\; \Gamma_{jk}(\psi^*) = \Gamma_i(\psi)\]
so double conjugation in the conjugate space is isomorphic to single conjugation in ordinary space.  It is straight forward to show: ($\bar{\psi}^{jk}=\psi^{\dag jk}\gamma_2\gamma_3 $) 
\begin{equation*}
\textcolor{red}{\partial_\mu(\bar{\psi}^{jk}\gamma_\mu\psi)= 0.} 
\end{equation*}
and likewise:
\begin{equation*}
\textcolor{green}{\partial_\mu(\bar{\psi}^{ki}\gamma_\mu\psi)= 0.}
\end{equation*}
\begin{equation*}
\textcolor{blue}{\partial_\mu(\bar{\psi}^{ij}\gamma_\mu\psi)= 0.}
\end{equation*}

\subsection{Triple (Complete) Quaternion Conjugation}
By taking the transpose and complete conjugation of ($\ref{eqn:dw}$) we get
\begin{equation}
\partial_\mu\psi^\dag\gamma_\mu^\dag - m\psi^\dag = 0 \label{eqn:d2}
\end{equation}
and since
\[\gamma_\mu^\dag = (-\gamma_0, \gamma_1, \gamma_2, \gamma_3) \]
and by multiplying on the right by $\gamma_0$ gives
\[\gamma_\mu^\dag\gamma_0 = \gamma_0(-\gamma_0, -\gamma_1, -\gamma_2, -\gamma_3) =-\gamma_0\gamma_\mu \]
($\ref{eqn:d2}$) becomes
\begin{equation*}
\partial_\mu\psi^\dag\gamma_0\gamma_\mu +m\psi^\dag\gamma_0 = 0 
\end{equation*}
and by setting
\[\psi^\dag\gamma_0 = \bar{\psi}\] we get
\begin{equation}
\partial_\mu\bar{\psi}\gamma_\mu + m\bar{\psi} = 0 \label{eqn:d3}
\end{equation}
Now by multiplying ($\ref{eqn:dw}$) on the left by $\bar{\psi}$ and ($\ref{eqn:d3}$) on the right by $\psi$ and adding we get
\[\partial_\mu\bar{\psi}\gamma_\mu\psi + \bar{\psi}\gamma_\mu\partial_\mu\psi = 0 \] 
or
\[\partial_\mu(\bar{\psi}\gamma_\mu\psi) = 0.\]
Thus in QM$_H$ three continuity equations - representing conservation of "color" - arise via a single conjugation (of each of $i,j,k$) or if a triple conjugation is used a continuity equation representing probability is obtained as it does in QM$_C$.  
\section{Speculation}
If, indeed, QM$_H$ is the appropriate mathematical approach to the strong force and QM$_C$ is appropriate for the electroweak force, might QM$_R$ be appropriate for gravity?  Apparently the particles decribed by QM$_H$ and QM$_C$ are subject to gravity and $\mathbb{C}$ and $\mathbb{H}$ each have $\mathbb{R}$ as a subfield.  Likewise, if QM$_C$ describes the electroweak force, and since $\mathbb{H}$ has $\mathbb{C}$ as a subfield the particles described in QM$_H$ should be (and are) subject to electroweak forces.  Perhaps the "real"s of $\mathbb{R},\; \mathbb{C}, \;$ and $\mathbb{H}$ are not the one and same but are independent with some means of interacting.  In a similar manner, perhaps the imaginary, $i$, of $\mathbb{C}$ and $\bar{q}$, (when treated as a constant with fixed "phase"), of $\mathbb{H}$, while isomorphic, are not the same and have a mechanism to interact.  Finally the imaginaries of $\mathbb{H}$, when treated as having a variable phase, are not isomorphic to a subfield in either $\mathbb{C}$ nor $\mathbb{R}$ and thus QM$_R$ and QM$_C$ can not describe particles that "strongly" interact.  Thus, perhaps,
\[\psi_{reality} = \psi_\mathbb{R} \cdot \psi_\mathbb{C} \cdot \psi_\mathbb{H}.\]
\newpage
\noindent{
\normalsize{\textbf{Appendix A, Quaternions}}}
\newline
\newline
Quaternions, $\mathbb{H}$, are one of only three ($\mathbb{R}$, $\mathbb{C}$ and $\mathbb{H}$) finite-dimensional division rings containing the real numbers  $\mathbb{R}$ as a subring - a requirement to preserve probability in quantum mechanics. $\mathbb{H}$ can be loosely viewed as a non-commutative extension of $\mathbb{C}$.  The imaginary quaternion units, i, j, k are defined by
\[ii = jj = kk = -1\] 
\[ij = -ji = k,\hspace{.3in}ki = -ik = j,\hspace{.3in}jk = - kj = i\]
A general quaternion \emph{q} can be written
\[ q = a + bi + cj + dk\]
where \[\emph{a,b,c,d} \in \mathbb{R}.\]
Every non-zero quaternion has an inverse.  
Quaternion addition is associative  - $q_1 + (q_2 + q_3) = (q_1 + q_2) + q_3$ - and defined as 
\[q_1 + q_2 = a_1 + a_2 +(b_1 + b_2)i + (c_1 +c_2)j + (d_1 +d_2)k\]
and quaternion multiplication (paying heed to the non-commutative nature of the imaginary units) is
\[q_1q_2 = (a_1a_2 - b_1b_2 - c_1c_2 - d_1d_2) \]
\[+(a_1b_2 + b_1a_2 + c_1d_2 - d_1c_2)i \]
\[+(a_1c_2 - b_1d_2 + c_1a_2 + d_1b_2)j \]
\[+(a_1d_2 + b_1c_2 - c_1b_2 + d_1a_2)k \]
Quaternions are associative under multiplication  $(q_1q_2)q_3 = q_1(q_2q_3)$.
A unit imaginary quaternion $\bar{q}$ is defined as
\[\bar{q} = b i + c j + d k\]
\begin{center} ( b, c, d $\in$ $\mathbb{R}$ )
\end{center}
where $\bar{q}^2 = -1$, which means $b^2 + c^2 + d^2 = 1.$
It should also be noted that if \emph{i} refers to the \emph{i} of $\mathbb{C}$ rather than of $\mathbb{H}$ it will be specifically indicated.
\newpage
\noindent{
\normalsize{\textbf{Appendix B, Dirac's Matrices, $\gamma^C$, in $\mathbb{C}$}}}
\vspace*{.1in}
\newline
One representation to satisfy ($\ref{eqn:dw}$) is 

\vspace*{.3in}
\large
$\gamma_x =
\left( \begin{array}{cccc}
 0 & 0 & 0 & -i\\
 0 & 0 & -i & 0\\
 0 & i & 0 & 0\\
 i & 0 & 0 & 0
      \end{array} \right) 
\equiv 
      \left( \begin{array}{cc}
      0 & -i\mathbf{\sigma_x} \\
      i\mathbf{\sigma_x} & 0
              \end{array}\right); \hspace{.2in} 
      \sigma_x=
           \left( \begin{array}{cc}
      0 & 1\\
      1 & 0
              \end{array}\right)$
\vspace*{.3in}

$\gamma_y =
\left( \begin{array}{cccc}
0 & 0 & 0 & -1\\
0 & 0 & 1 & 0\\
0 & 1 & 0 & 0\\
-1 & 0 & 0 & 0
      \end{array} \right)
 \equiv
            \left( \begin{array}{cc}
      0 & -i\mathbf{\sigma_y} \\
      i\mathbf{\sigma_y} & 0
              \end{array}\right); \hspace{.2in} 
       \sigma_y=
           \left( \begin{array}{cc}
      0 & -i\\
      i & 0
              \end{array}\right)$             
\vspace*{.3in}

$\gamma_z = 
\left( \begin{array}{cccccccc}
0 & 0 & -i & 0\\
0 & 0 & 0 & i\\
i & 0 & 0 & 0\\
0 & -i & 0 & 0
       \end{array} \right)
       \equiv
                   \left( \begin{array}{cc}
      0 & -i\mathbf{\sigma_z} \\
      i\mathbf{\sigma_z} & 0
              \end{array}\right);\hspace{.2in}
        \sigma_z=
           \left( \begin{array}{cc}
      1 & 0\\
      0 & -1
              \end{array}\right)$              
\vspace*{.3in}

$\gamma_t =
\left( \begin{array}{cccccccc}
i & 0 & 0 & 0 \\
0 & i & 0 & 0 \\
0 & 0 & -i & 0\\
0 & 0 & 0 & -i 
\end{array} \right)
 \equiv
          \left( \begin{array}{cc}
      \mathbf{\sigma_t} & 0 \\
      0 & -\mathbf{\sigma_t}
              \end{array}\right);\hspace{.2in}
        \sigma_t=
           \left( \begin{array}{cc}
      i & 0\\
      0 & i
              \end{array}\right)$ 
              
\vspace*{.3in}

Note $\sigma_x \sigma_y = i\sigma_z=\mathbf{I}_C\sigma_z$ cyclically, are rank 2, and $\in \mathbb{C}$ and $\sigma_t = \mathbf{I}_C \otimes \mathbf{1}_2 .$
\newpage
\noindent{
\normalsize{\textbf{Appendix C, Dirac's Matrices in $\mathbb{R}$}}}
\vspace*{.5in}
\newline
One dimension 8 representation to satisfy ($\ref{eqn:dw}$) is 

\vspace*{.3in}

\large
$\gamma_x =$
\tiny
$\left( \begin{array}{cccccccc}
0 & 0 & 0 & 0 & 0 & 0 & 0 & -1\\
0 & 0 & 0 & 0 & 0 & 0 & 1 & 0\\
0 & 0 & 0 & 0 & 0 & -1 & 0 & 0\\
0 & 0 & 0 & 0 & 1 & 0 & 0 & 0\\
0 & 0 & 0 & 1 & 0 & 0 & 0 & 0\\
0 & 0 & -1 & 0 & 0 & 0 & 0 & 0\\
0 & 1 & 0 & 0 & 0 & 0 & 0 & 0\\
-1 & 0 & 0 & 0 & 0 & 0 & 0 & 0
      \end{array} \right)$ 
\small
$\equiv 
      \left( \begin{array}{cc}
      0 & \sigma_x \\
      -\sigma_x & 0
              \end{array}\right);\hspace{.2in}
       \sigma_x=$\tiny
       $\left( \begin{array}{cccc}
      0 & 0 & 0 & -1 \\
      0 & 0 & 1 & 0\\
      0 & -1& 0 & 0\\
      1 & 0 & 0 & 0
              \end{array}\right)$
              
\vspace*{.3in}
\large
$\gamma_y =$
\tiny
$\left( \begin{array}{cccccccc}
0 & 0 & 0 & 0 & 0 & 0 & -1 & 0\\
0 & 0 & 0 & 0 & 0 & 0 & 0 & -1\\
0 & 0 & 0 & 0 & 1 & 0 & 0 & 0\\
0 & 0 & 0 & 0 & 0 & 1 & 0 & 0\\
0 & 0 & 1 & 0 & 0 & 0 & 0 & 0\\
0 & 0 & 0 & 1 & 0 & 0 & 0 & 0\\
-1 & 0 & 0 & 0 & 0 & 0 & 0 & 0\\
0 & -1 & 0 & 0 & 0 & 0 & 0 & 0
      \end{array} \right)$
\small
$\equiv
     \left( \begin{array}{cc}
      0 & \sigma_y \\
      -\sigma_y & 0
              \end{array}\right);\hspace{.2in}
                    \sigma_y=$\tiny
       $\left( \begin{array}{cccc}
      0 & 0 & -1 & 0 \\
      0 & 0 & 0 & -1\\
      1 & 0& 0 & 0\\
      0 & 1 & 0 & 0
              \end{array}\right)$ 
              
\vspace*{.3in}
\large
$\gamma_z = $
\tiny
$\left( \begin{array}{cccccccc}
0 & 0 & 0 & 0 & 0 & -1 & 0 & 0\\
0 & 0 & 0 & 0 & 1 & 0 & 0 & 0\\
0 & 0 & 0 & 0 & 0 & 0 & 0 & 1\\
0 & 0 & 0 & 0 & 0 & 0 & -1 & 0\\
0 & 1 & 0 & 0 & 0 & 0 & 0 & 0\\
-1 & 0 & 0 & 0 & 0 & 0 & 0 & 0\\
0 & 0 & 0 & -1 & 0 & 0 & 0 & 0\\
0 & 0 & 1 & 0 & 0 & 0 & 0 & 0
       \end{array} \right)$
\small       
$\equiv
     \left( \begin{array}{cc}
      0 & \sigma_z \\
      -\sigma_z & 0
              \end{array}\right);
              \hspace{.2in}
      \sigma_z=$\tiny
       $\left( \begin{array}{cccc}
      0 & -1 & 0 & 0 \\
      1 & 0 & 0 & 0\\
      0 & 0& 0 & 1\\
      0 & 0 & -1 & 0
              \end{array}\right)$ 
\vspace*{.3in}

\large
$\gamma_t=$
\tiny
$\left( \begin{array}{cccccccc}
0 & 1 & 0 & 0 & 0 & 0 & 0 & 0\\
-1 & 0 & 0 & 0 & 0 & 0 & 0 & 0\\
0 & 0 & 0 & 1 & 0 & 0 & 0 & 0\\
0 & 0 & -1 & 0 & 0 & 0 & 0 & 0\\
0 & 0 & 0 & 0 & 0 & -1 & 0 & 0\\
0 & 0 & 0 & 0 & 1 & 0 & 0 & 0\\
0 & 0 & 0 & 0 & 0 & 0 & 0 & -1\\
0 & 0 & 0 & 0 & 0 & 0 & 1 & 0
\end{array} \right)$
\small
$ \equiv
          \left( \begin{array}{cc}
      \sigma_t & 0 \\
      0 & -\sigma_t
              \end{array}\right);
              \hspace{.2in}
                    \sigma_t=$\tiny
       $\left( \begin{array}{cccc}
      0 & 1 & 0 & 0 \\
      -1 & 0 & 0 & 0\\
      0 & 0& 0 & 1\\
      0 & 0 & -1 & 0
              \end{array}\right)$  
\vspace*{.3in}
\large

Note: $\sigma_x \sigma_y = \sigma_z$ cyclically, are rank 4, and $\in \mathbb{R}$ and $\sigma_t = \mathbf{I}_R \otimes \mathbf{1}_2 .$
\newpage
\vspace*{.5in}
\noindent{
\normalsize{\textbf{Appendix D, Dirac's (4x4) Matrices, $\gamma^R$ in $\mathbb{R}$}}}
\vspace*{.5in}
\newline
One dimension 4 representation to satisfy ($\ref{eqn:dw}$) is \\

$\gamma_x =$
$\left( \begin{array}{cccc}
1 & 0 & 0 & 0 \\
0 & 1 & 0 & 0 \\
0 & 0 & -1 & 0 \\
0 & 0 & 0 & -1 \\
      \end{array} \right)$ 
$\equiv 
      \left( \begin{array}{cc}
      \sigma_x & 0\\
      0& -\sigma_x
              \end{array}\right);\hspace{.2in}
       \sigma_x=
       \left( \begin{array}{cc}
      1 & 0 \\
      0 & 1\\
             \end{array}\right)$
\vspace*{.3in}

$\gamma_y =$
$\left( \begin{array}{cccc}
0 & 0 & 1 & 0 \\
0 & 0 & 0 & 1 \\
1 & 0 & 0 & 0 \\
0 & 1 & 0 & 0 \\
      \end{array} \right)$
$\equiv
     \left( \begin{array}{cc}
      0 & \sigma_y \\
      -\sigma_y & 0
              \end{array}\right);\hspace{.2in}
                    \sigma_y=
       \left( \begin{array}{cc}
      1 & 0 \\
      0 & 1 \\
           \end{array}\right)$ 
              
\vspace*{.3in}
$\gamma_z = $
$\left( \begin{array}{cccc}
0 & 0 & 0 & 1\\
0 & 0 & -1 & 0 \\
0 & -1 & 0 & 0\\
1 & 0 & 0 & 0\\
       \end{array} \right)$
$\equiv
     \left( \begin{array}{cc}
      0 & \sigma_z \\
      -\sigma_z & 0
              \end{array}\right);
              \hspace{.2in}
      \sigma_z=
      \left( \begin{array}{cccc}
      0 & 1 \\
      -1 & 0\\
            \end{array}\right)$ 
            
\vspace*{.3in}
$\gamma_{t}=$
$\left( \begin{array}{cccc}
0 & 0 & 0 & 1 \\
0 & 0 & 1 & 0\\
0 & -1 & 0 & 0 \\
-1 & 0 & 0 & 0 \\
\end{array} \right)$
$ \equiv
          \left( \begin{array}{cc}
      0 &\sigma_t  \\
      -\sigma_t & 0
              \end{array}\right);
              \hspace{.2in}
                    \sigma_t=
       \left( \begin{array}{cc}
      0 & 1 \\
      1 & 0 \\
           \end{array}\right)$

\vspace*{.3in}
\newpage
\noindent{
\normalsize{\textbf{Appendix E, Dirac's Matrices in $\mathbb{H}$}}}
\vspace*{.5in}
\newline
One representation, maintaining $\emph{i,j,k}$ symmetry, satisfying ($\ref{eqn:dw}$) is 

\vspace*{.3in}

\large
$\gamma_x =
\left( \begin{array}{cccc}
 0 & i & 0 & 0\\
 -i & 0 & 0 & 0\\
 0 & 0 & 0 & i\\
 0 & 0 & -i & 0
      \end{array} \right) 
\equiv
i\left( \begin{array}{cc}
      \mathbf{\sigma} & 0 \\
      0 &\mathbf{\sigma}
      \end{array}\right);\hspace{.2in}
        \sigma=
   \left( \begin{array}{cc}
      0 & 1\\
      -1 & 0
          \end{array}\right)$
\vspace*{.3in}

$\gamma_y =
\left( \begin{array}{cccc}
0 & j & 0 & 0\\
-j & 0 & 0 & 0\\
0 & 0 & 0 & j\\
0 & 0 & -j & 0
      \end{array} \right)
\equiv
j\left( \begin{array}{cc}
      \mathbf{\sigma} & 0 \\
      0 &\mathbf{\sigma}
      \end{array}\right)$             
\vspace*{.3in}

$\gamma_z = 
\left( \begin{array}{cccccccc}
0 & k & 0 & 0\\
-k & 0 & 0 & 0\\
i & 0 & 0 & k\\
0 & 0 & -k & 0
       \end{array} \right)
 \equiv
k\left( \begin{array}{cc}
      \mathbf{\sigma} & 0 \\
      0 &\mathbf{\sigma}
      \end{array}\right)$
\vspace*{.3in}

$\gamma_t =
\left( \begin{array}{cccccccc}
0 & 0 & -1 & 0 \\
0 & 0 & 0 & 1 \\
1 & 0 & 0 & 0\\
0 & -1 & 0 & 0 
\end{array} \right)
\equiv
\left( \begin{array}{cc}
      0& -\mathbf{\sigma}_t  \\
      \mathbf{\sigma}_t& 0
      \end{array}\right);\hspace{.2in}
        \sigma_t=
   \left( \begin{array}{cc}
      1 & 0\\
      0 & -1
          \end{array}\right).$ 
\newpage
\bibliography{quaternion}
\bibliographystyle{unsrt}

\end{document}